\documentclass[usenatbib,usegraphicx,twocolumn]{mn2e}

\newcommand{\be}{\begin{equation}}
\newcommand{\e}{\end{equation}}
\newcommand{\bear}{\begin{eqnarray}}
\newcommand{\ear}{\end{eqnarray}}

\def\apj{ApJ}

\def\mnras{MNRAS}

\def\aap{A\&A}
\def\apjs{ApJS}

\begin{document}
\title[The  Local Dimension]{The Local Dimension: a method to quantify
  the Cosmic Web} 

\author[Sarkar \& Bharadwaj]
{Prakash Sarkar\thanks{E-mail: prakash@cts.iitkgp.ernet.in}
and
Somnath Bharadwaj\thanks{E-mail: somnathb@iitkgp.ac.in}\\
Department of Physics \& Meteorology\\
\hspace{.5in} and\\
Centre for Theoretical Studies,\\
 Indian Institute of Technology,\\
 Kharagpur 721302, India\\ }

\maketitle

\date{\today}

\begin{abstract}
It is now well accepted that the galaxies are distributed in 
filaments, sheets and  clusters all of which form  an 
interconnected network known as the Cosmic Web. It is a big  challenge
to quantify the shapes of the  interconnected structural
elements  that  form this   network. Tools  like the Minkowski
functionals  which use  global properties, though
well suited for an isolated object like  a single sheet or filament,
are not suited for an interconnected network of such objects.   
 We consider the Local Dimension $D$, defined through $N(R)=A R^D$,
 where $N(R)$ is the galaxy number count within a sphere of comoving
 radius  $R$ centered on  a particular galaxy, as a tool to locally
 quantify  the shape  in the neigbourhood of different galaxies along
 the Cosmic Web. We expect $D \sim 1,2$ and $3$ for a galaxy located
 in a filament, sheet and cluster respectively. Using 
LCDM  N-body simulations  we find  that it is possible to
determine $D$ through a power law fit to $N(R)$ across the
length-scales $2$ to $10 \, {\rm Mpc}$ for $\sim 33 \%$ of the
galaxies. We have visually identified the filaments and sheets
corresponding to many of the galaxies  with $D \sim 1$ and $2$
respectively. In several other situations the structure responsible
for the $D$  value could not be visually identified, either due to its 
being  tenuous or due to  other dominating 
structures in the vicinity. We also show that  the global
distribution of the $D$ values can be used to visualize and interpret 
 how the different structural elements are woven into the Cosmic Web.
\end{abstract}


\section{Introduction}
Filaments are  the most prominent  features visible in the galaxy 
distribution.  This finding dates back to a few papers in the 
seventies and eighties \citep{joe, einas4,zel}. Subsequent work 
substantiates this 
(e.g. \citealt{gel}, \citealt{shect}, \citealt{shand2},
 \citealt{bharadwaj2000}, \citealt{mul}, \citealt{basil}, \citealt{doro2},
 \citealt{pimb}) and shows the filaments to be statistically
 significant \citep{bharadwaj2004,pandey}. 
It is now well accepted that galaxies are distributed in an
interconnected network of clusters, sheets and filaments
encircling voids. This complicated  pattern is often referred to as the
Cosmic Web.  Despite this progress, it still remains a challenge to
quantify the Cosmic Web that is so distinctly visible in 
galaxy redshift surveys (eg. SDSS DR5, \citet{adel}).

Statistical measures like  the  void probability function
\citep{White1979}, percolation analysis \citep{shandarin1983} and  the
genus curve \citep{Gott1} each quantifies a different aspect of the
Cosmic Web. The Minkowski functionals \citep{mecke1994} are very
effective to quantify the shapes of  individual structural elements
like sheets or filaments. In $3$ dimensions there are $4$ Minkowski
functionals, namely the volume, surface area, integrated mean
curvature and integrated Gaussian curvature.  \citet{Sahni1998} 
introduce   the 'Shapefinders', essentially ratios of the  Minkowski
functionals,  as a very effective shape diagnostic. A $2$ dimensional
version of Shapefinders \citep{bharadwaj2000} has been  extensively
used  to quantify  the filamentarity in the galaxy distribution 
(\citealt{pandey3} and references therein).

\begin{figure}
\includegraphics[width=0.4\textwidth]{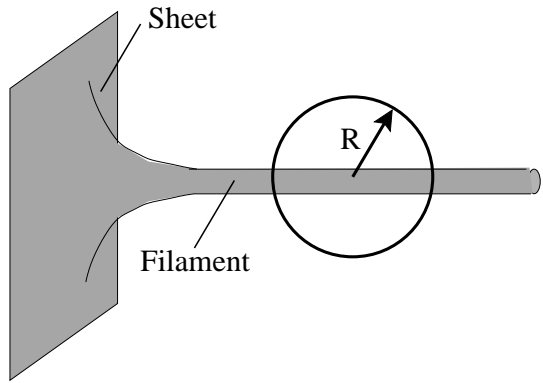}
\caption{This shows an interconnected  sheet and filament. Also shown, 
a sphere of  radius $R$ centered on a galaxy located in the filament.} 
\label{fig:exp1}
\end{figure}

Though the Minkowski functionals and the Shapefinders are very
effective techniques to quantify the shapes of individual structural
elements like sheets or filaments,  it is very  different when
dealing with the  Cosmic Web which is an interconnected network of 
filaments, sheets and clusters. For example consider a sheet
connected to a filament as shown in Figure \ref{fig:exp1}.  The
Minkowski functionals are global properties of the entire object {\it
  ie.} the area is the sum of the areas of the sheet and the
filament etc., and the fact that  object is actually a combination of
two different elements would be lost. It is necessary to quantify the
local shape at different points in the object in order to determine
that it actually is a combination of a sheet and a filament. 
In this paper  we consider the ``Local Dimension'' as a means to
quantify the local shape of the galaxy distribution at different
positions along the Cosmic Web. 

\section{Method of Analysis}

We choose  a particular galaxy as  center and  determine 
$N(R)$ the number of other galaxies within   a  sphere 
of comoving radius $R$. This is done varying
$R$. In the situation where a power law
\begin{equation}
N(R)=A R^D
\label{eq:1}
\end{equation}
gives a good fit over the  length-scales  $R_1 \le R \le R_2$, 
we identify $D$ as the Local Dimension in the neighbourhood of the
center. The values $D=1, 2$ and $3$ correspond to a filament, sheet
and cluster respectively. It may be noted that the term ``cluster'' 
here denotes a three dimensional, volume filling structural element
and is not to be confused with a ``cluster of galaxies''.  
Values of $D$ other
 than $1,2$ and $3$ are more difficult to  interpret. For example, 
a galaxy  distribution that is more diffuse  than a filament but does
not fill a plane would give a fractional value (fractal) in the range
$1 \le D \le 2$. 
 Referring to Figure \ref{fig:exp1}, we expect
$D=1$ and $D=2$ when the center is located in the filament and  the
sheet respectively. This is provided that the center is well away from the
intersection of  the filament and  the sheet. 
When the intersection lies within   $R_1 \le R \le R_2$ from the 
 center, there will be a change in the slope of $N(R)$ when it crosses
 the intersection.  It is not possible to determine a local dimension
 at  the centers where such a  situation occurs.

 We  perform  this analysis using every  galaxy in the sample 
 as a center. In general it will be possible to determine a Local
 Dimension for only  a fraction of the galaxies. 
It is expected that  with a suitable  choice of
the $R$ range {\it ie.}  $R_1$ and $R_2$,  it will be possible to 
determine the Local Dimension for   a  substantial number of the
centers. The value of the Local Dimension at different positions will
 indicate the location of the filaments, sheets and clusters and
 reveal how these are woven into the Cosmic Web. 

\begin{figure}
\rotatebox{270}{\includegraphics[height=0.4\textwidth]{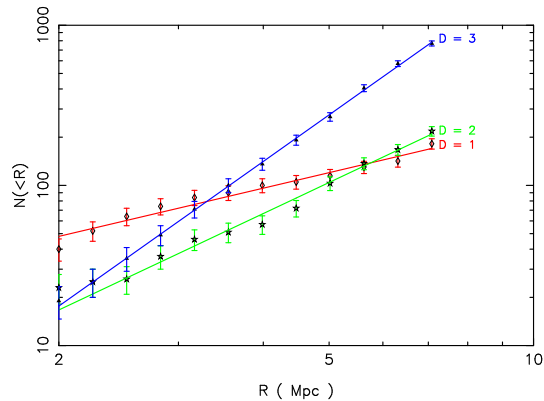}}
\caption{This shows $N(R)$ as a function of $R$ for three different
  centers from a particular realization. The centers have been chosen
  so as to demonstrate power law fits with $D \sim 1,2$ and $3$
  corresponding to a 
  filament, sheet and cluster respectively.} 
\label{fig:exp2}
\end{figure}

In this {\em Letter} we test this idea and demonstrate its utility by
applying it to simulations.  We have used  a Particle-Mesh (PM) 
N-body code to simulate the $z=0$  dark matter distribution. The
simulations have $256^3$ particles on a  $256^3$ mesh with  grid
spacing  $0.5  \, {\rm Mpc} $. The simulations were carried
out using a LCDM power spectrum with the  parameters
$(\Omega_{m0}, \Omega_{\Lambda0}, h, n_s, \sigma_8)=(0.3, 0.7, 0.7, 1,1)$. 
We have identified $210,000$ particles, randomly  drawn from the simulation
output, as galaxies.  These have a  mean interparticle separation of
$\sim 2 ~ {\rm Mpc}$, comparable to that in galaxy surveys.   This
simulated galaxy distribution was carried over to redshift space   
in the plane parallel approximation. The subsequent analysis to
determine the Local Dimension was carried out using this simulated sample of
$210,000$ galaxies. Since the resolution of the simulation is about
$1\, {\rm Mpc}$,  we can't choose  $R_1$ to be less than
that. The value of $R_2$ is determined by the limited box size.   
We have chosen the value of $R_1$ and $R_2$ to be
 $2$ and $10  \, {\rm  Mpc }$ respectively. Increasing $R_2$ causes a
considerable drop in the number of centers for which the  Local
Dimension is defined. 

The analysis was carried out for $10$ different, independent
realizations of the dark matter distribution. Figure \ref{fig:exp2}
shows $N(R)$ for three different centers chosen from a particular
realization.  The $1-\sigma$ error at each data point is
$\sqrt{N(r)}$ due to the Poisson fluctuation. For each center we have
  determined the power law $N(R)=A R^D$ that provides the best fit  to the
data. The power law fit is accepted if the chi-square per degree of
freedom satisfies $\chi^2/\nu \le 1.2$ and  the value of $D$ is
accepted as the Local Dimension corresponding to the particular
center. The power law fit is rejected for larger values of
$\chi^2/\nu$, and  the Local Dimension is undetermined for the
particular center.   
 The criteria  $\chi^2/\nu \le 1.2$ is chosen as a   compromise between
 a good power 
law fit and a reasonably large number of centers for which $D$ can 
been determined.  The number of centers for which  $D$ can  been
determined falls if a more stringent criteria is imposed on
$\chi^2/\nu$. 

\section{Results and Conclusions}

\begin{figure}
\includegraphics[width=.45\textwidth]{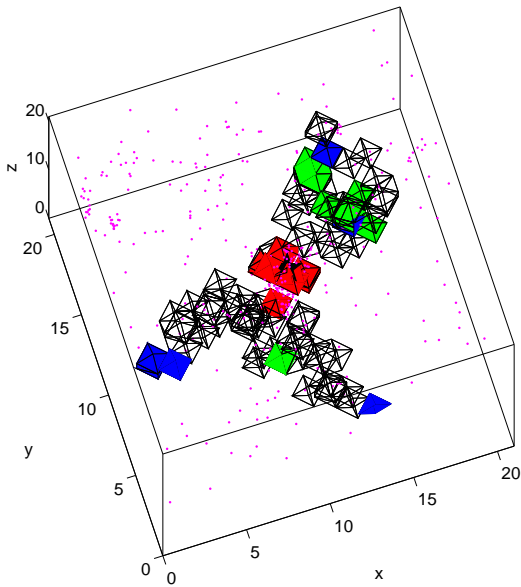}
\caption{The  galaxies within $10 \, {\rm Mpc}$ of a  center with 
 $D \sim 1$ are shown as points. The galaxy distribution was
 converted to a set of 1s and 0s on a grid of spacing $1 \, {\rm
 Mpc}$. Connected structures were identified using 
 the Friend-of-Friend algorithm where adjacent 1s are identified
 as belonging to the same connected structure.
Only the two largest structures have been 
 shown, both pass very close to the center and they appear connected
 in the figure.  The $D$ values have been painted on these structures
 with red, blue and green representing $D \sim 1,2$ and $3$
 respectively. The $D$ value is undetermined for  the  parts of the
 structures  shown in black.} 
\label{fig:fil1n}
\end{figure}

The number of centers for which it is
possible to determine a Local Dimension varies between $69,017$ to
$73,481$ {\it ie.} less than $7 \%$ variation across the $10$
different realizations. The $D$ values in the interval
$1 \pm 0.5$, $2 \pm 0.5$ and $3 \pm 0.5$ have respectively been binned
as $D \sim 1,2$ and $3$. We  have  inspected  the galaxy
distribution in the vicinity of a few  centers in order to visually 
identify the structures  corresponding to the respective $D$ values. 
Figure \ref{fig:fil1n} shows the galaxy distribution 
within $10 \, {\rm Mpc}$ of a center which has Local Dimension $ D
\sim 1$.  We expect this center to be located in a filament. 
We have used  the Friend-of-Friend algorithm with linking
length $\sqrt{3}  \, {\rm Mpc}$  to identify   connected patterns in
the galaxy distribution. We note that all the galaxies in this region
connect up into a single structure if the linking length is doubled to
$2 \times \sqrt{3} \, {\rm Mpc}$.  Using $\sqrt{3}  \, {\rm Mpc}$ 
yields several  disconnected  structures of which  
we show only the  largest two.  Both the structures 
 pass  close to the center, and appear connected in the figure. Of the
 two structures,  
one  with two  well separated  tentacles appears below the center,
whereas the other 
with two 
nearly connected tentacles appears  above it. 
 The filamentary  nature of these structures is quite evident. 
Out of the total $79$ galaxies in these two structures, $D$ is
undetermined for $58$. There are $8$ galaxies with $D\sim 1$, all 
located in a contiguous region near the center,  and   $6$ and $7$ galaxies 
with $D\sim 2$ and $3$ respectively spread out along the
structure  at 
locations quite different  from the center. While it is relatively
easy  to visually identify the  filament corresponding to the centers
with  $D\sim1$, we have not been able to visually identify the sheets and
clusters corresponding to  the centers with $D\sim 2$ and $3$ shown
in   Figure \ref{fig:fil1n}. We note that a  visual identification of
the sheets and clusters  corresponding to $D \sim 2$ and $3$
respectively has been  possible in other fields where the largest
structure is predominantly sheetlike or clusterlike.  It is quite
apparent that the largest structure in Figure \ref{fig:fil1n} is
predominantly filamentary.

\begin{figure}
\includegraphics[width=.45\textwidth]{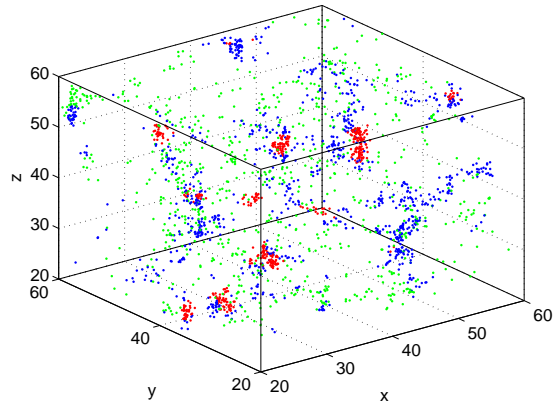}
\caption{This  shows the distribution of $D$ values in a box of size
$[40 \, {\rm Mpc}]^3$,  red, blue and green denote 
  $D \sim 1,2$ and $3$ respectively.}
\label{fig:allshown}
\end{figure}

Figure \ref{fig:allshown} shows the  distribution of $D$ values
over a large   region from one of the realizations. The distribution
shows   
distinct patterns, there being 
regions with size ranging from a few
${\rm   Mpc}$    to tens of ${\rm Mpc}$ where the $D$ value is
constant. The centers  with $D \sim 1$ appear to have a more  dense
and  compact  distribution compared to the centers with $D \sim 2$, 
whereas the centers with $D \sim 3$ appear to  have a 
 rather diffuse distribution. 

\begin{figure}
\rotatebox{270}{\includegraphics[height=0.45\textwidth]{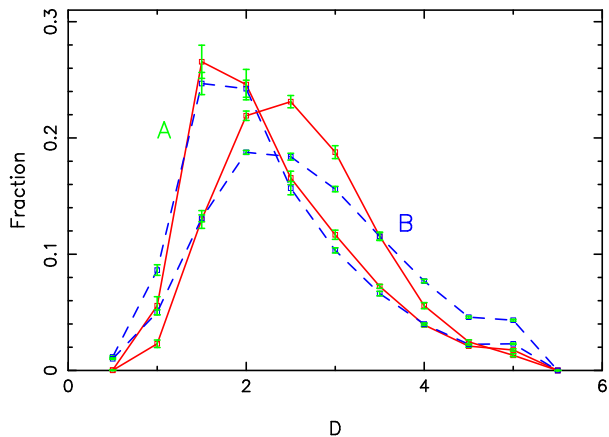}}
\caption{(A) The fraction of centers with a particular $D$ value, and
  (B) the volume fraction  containing a particular $D$ value. The bins
  in $D$ have size $\pm 0.25$. The solid and dashed curves use $D$
  values   determined over length-scales $2-10 \, {\rm Mpc}$ and 
$2-5 \, {\rm Mpc}$ respectively.
}
\label{fig:hist1}
\end{figure}

\begin{figure}
\rotatebox{270}{\includegraphics[height=.45\textwidth]{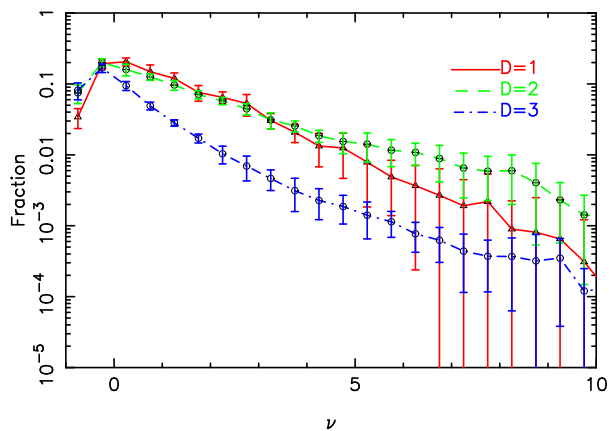}}
\caption{For each $D$ value we separately show  the fraction
  of centers  as a function of the  density environment. The latter is
  quantified through $\nu=\delta/\sigma$ where $\delta$ is the density
  contrast and $\sigma^2$ its  variance. This is evaluated on a grid
  of spacing $1 \,  {\rm Mpc}$  and the density field is smoothened
  with a Gaussian of width $5 \, {\rm Mpc}$.  
}  
\label{fig:hist2}
\end{figure}

Figure \ref{fig:hist1} separately quantifies two different  aspects of  
the distribution of  $D$ values, (A.) the fraction  of centers 
with a particular  $D$ value, and (B.)  the fraction  of volume  
with a particular  $D$ value. For this analysis 
the $D$ values  were divided into bins of width $\pm 0.25$ and 
the centers for which a $D$ value could not 
be determined were discarded. 
We find that the fraction of centers and the volume fraction both 
have  very robust statistics with a variation of $\sim 10
\%$ or less across the $10$ different realizations. 
We first consider the distribution of
the fraction  of centers with different  $D$ values. 
  The bin  at $D=1.5$ contains the
maximum number of centers, and the two  bins  at
$D=1.5$ and $2$ together contain more than $50 \%$ of the centers. 
This indicates that the maximum number   of centers for which $D$ can be
determined lie in  sheets and filaments. Assuming that this is
representative of the entire matter distribution, we conclude that the
bulk of the matter is contained in sheets and filaments, with the
sheets dominating. We next consider the
 the fraction of  volume occupied by  a particular $D$ value. 
We have estimated this using a mesh with grid spacing $1 \, {\rm
  Mpc}$.  Each  grid  position was assigned the $D$ value of the
centers within the corresponding  grid cell. Cells that contain 
centers with different   values of $D$ were discarded. 
It was possible to assign a $D$ value to $\sim 35,000$ grid cells.
We find that the distribution peaks at $D=2.5$, the peak 
is rather broad and more than $50 \%$ of the volume  for which it was
possible to assign a $D$ value  is occupied by  values in the range
$2$ to $3$. Assuming that this  result is  valid for the entire
volume, we conclude that the volume is predominantly filled with
sheets and clusters.  While the matter is mainly
contained in sheets and filaments, the filaments do not occupy a very
large fraction of the volume. This leads to a picture where the
filaments have relatively higher  matter  densities  as compared to  
clusters. To test this we  separately consider the centers with 
$D$ values $1,2$ and $3$, and for each value we determine the fraction
of centers as a function of the  density environment. The density was  
calculated on a $1 \, {\rm Mpc}$ grid after smoothing with  a
Gaussian filter of width $5 \, {\rm Mpc}$.  We find  (Figure
\ref{fig:hist2}) that in comparison  to the centers with $D \sim 1$
and $2$, the centers  with  $D \sim 3$  are more abundant  in under-dense
regions  in preference to the over-dense regions.  This is consistent
with the picture where  filaments and sheets located along the Cosmic
Web  contain  the bulk of the matter and also  most of the high
density regions, whereas the centers  with $D
\sim 3$  are predominantly located in under-dense regions away from the
Cosmic Web, namely the voids.  

The results presented above are specific to the length-scale range
$2-10 \, {\rm Mpc}$.  It is quite  possible that the relative
abundance  of clusters, sheets and filaments  would be quite
different if the same analysis were carried out over a considerably
different range of length-scales, say for example $0.1 - 1 \, {\rm
  Mpc}$.  Here we have repeated the entire analysis using a smaller
range of length-scales $2-5 \, {\rm Mpc}$ for which the results are
also shown in Figure \ref{fig:hist1}. We find that the results are
qualitatively similar to those obtained using  $2-10 \, {\rm Mpc}$,
indicating that our  conclusions regarding the relative abundance of
clusters, sheets and filaments  are quite robust and are not very
sensitive to small changes in the range over which $D$  is
determined.  This also indicates  that the $D$ values determined 
using length-scales $2-10 \,{\rm Mpc}$ are not severely biased by the
presence of other structural elements with $5-10 \, {\rm   Mpc}$
of the one on which the center is located.

In conclusion we note that the Local Dimension provides a robust
method to 
quantify the shapes and probe the distribution of the  different,
interconnected   structural elements that  make up the Cosmic Web. We
plan to apply this to the SDSS and other galaxy surveys in future. 

\section*{Acknowledgment}
P.S. is thankful to Biswajit Pandey, Kanan Datta and Prasun Dutta for
useful discussions. P.S. would like to acknowledge Senior Research
Fellow of University Grants Commission (UGC), India. for providing
financial support.


\begin{thebibliography}{99}


\bibitem[\protect\citeauthoryear{Adelman-McCarthy et al.}{2006}]{adel} 
Adelman-McCarthy, J.~K., et al.\ 2006, \apjs, 162, 38 


\bibitem[\protect\citeauthoryear{Basilakos, Plionis, \&
Rowan-Robinson}{2001}]{basil} Basilakos, S., Plionis, M., \&
Rowan-Robinson, M.\ 2001, \mnras, 323, 47

\bibitem[\protect\citeauthoryear{Bharadwaj et al.}{2004}]{bharadwaj2004}
Bharadwaj S., Bhavsar S. P., \& Sheth J. V., 2004, \apj, \  606, 25

\bibitem[\protect\citeauthoryear{Bharadwaj et al.}{2000}]{bharadwaj2000}
Bharadwaj S., Sahni V., Satyaprakash B. S.,Shandarin S. F., \& Yess C., 2000, \apj, \  528, 21



\bibitem[\protect\citeauthoryear{Doroshkevich et al.}{2004}]{doro2}
Doroshkevich, A., Tucker, D.~L., Allam, S., \& Way, M.~J. \ 2004, \aap, 
418, 7


\bibitem[\protect\citeauthoryear{Einasto et al.}{1980}]{einas4}
Einasto, J., Joeveer, M., \& Saar, E.\ 1980, \mnras, 193, 353


\bibitem[\protect\citeauthoryear{Geller \& Huchra}{1989}]{gel}
Geller, M.J. \& Huchra, J.P.1989, Science, 246, 897

\bibitem[\protect\citeauthoryear{Gott, Mellot \& Dickinson}{1986}]{Gott1}
Gott J. R., Mellot, A. L., \& Dickinson, M. \ 1986, \apj, 306, 341

\bibitem[\protect\citeauthoryear{Joeveer et al.}{1978}]{joe} Joeveer,
M., Einasto, J., \& Tago, E.\ 1978, \mnras, 185, 357

\bibitem[\protect\citeauthoryear{Mecke et al.}{1994}]{mecke1994}
Mecke K. R., Buchert T. \& Wagner H., 1994, A\&A, 288, 697

\bibitem[\protect\citeauthoryear{ M{\"u}ller et al.}{2000}] {mul} 
M{\"u}ller, V., Arbabi-Bidgoli, S.,  Einasto, J., \&
Tucker, D.\ 2000, \mnras, 318, 280  


\bibitem[\protect\citeauthoryear{Pandey \& Bharadwaj}{2005}]{pandey}
 Pandey, B. \& Bharadwaj, S. \ 2005, \mnras, 357, 1068

\bibitem[\protect\citeauthoryear{Pandey \& Bharadwaj}{2008}]{pandey3}
 Pandey, B. \& Bharadwaj, S. \ 2008, \mnras, In Press


\bibitem[\protect\citeauthoryear{Pimbblet, Drinkwater \& Hawkrigg}{2004}]{pimb}
Pimbblet, K. A., Drinkwater, M. J., \& Hawkrigg, M. C. \ 2004, \mnras,
354, L61 

\bibitem[\protect\citeauthoryear{Sahni et al.}{1998}]{Sahni1998}
Sahni V., Satyaprakash B. S., \& Shandarin S. F., 1998, \apj, 495, L5

\bibitem[\protect\citeauthoryear{Shandarin \& Yess}{1998}]{shand2} 
Shandarin, S.~F.~\& Yess, C.\ 1998, \apj, 505, 12 

\bibitem[\protect\citeauthoryear{Shandarin \& Zeldovich}{1983}]
{shandarin1983}Shandarin S. F., \& Zeldovich I. B., \  1983, Comments on 
Astrophysics, 10, 33

\bibitem[\protect\citeauthoryear{Shectman et al.}{1996}]{shect} 
Shectman, S.~A.,Landy, S.~D., Oemler, A., Tucker, D.~L., Lin, H.,
Kirshner, R.~P., \& Schechter, P.~L.\ 1996, \apj, 470, 172 

\bibitem[\protect\citeauthoryear{White}{1979}]{White1979}
White S. D. M., \ 1979 \mnras, 186, 145   

\bibitem[\protect\citeauthoryear{Zel'dovich , Einasto \&
    Shandarin}{1982}]{zel} Zel'dovich, I.~B., Einasto, J., \&
    Shandarin, S.~F.\  1982, Nature, 300, 407





\end{thebibliography}
\end{document}